# Sea Foam contains Hemoglycin from Cosmic Dust.

Julie E. M. McGeoch[1] and Malcolm W. McGeoch[2]

[1] High Energy Physics DIV, Smithsonian Astrophysical Observatory Center for Astrophysics | Harvard & Smithsonian, 60 Garden Str, MS 70, Cambridge MA 02138, USA
[2] PLEX Corporation, 275 Martine Str, Suite 100, Fall River, MA 02723, USA

**Abstract:**
In-falling cosmic dust has left evidence of meteoritic polymer amide in stromatolites, both fossil and modern. In search of evidence for continued present day in-fall sea foam was collected from two beaches in Rhode Island and subjected to Folch extraction to concentrate amphiphilic components in a chloroform water-methanol interphase layer. Hemoglycin polymer amide molecules previously characterized by MALDI mass spectrometry in meteorites and stromatolites were identified in sea foam either directly, or via their fragmentation patterns. Residual isotope enrichment pointed to an extra-terrestrial origin. The unique resiliency of sea foam may be due to the formation of extended hemoglycin lattices that stabilize its closed-cell structure and its lightness can potentially be explained by photolytic hydrogen production.

**Introduction**
Sea foam, a closed-cell foam that accumulates on shorelines world-wide was investigated to determine whether it contained hemoglycin, a polymer of glycine and iron present in dust arriving from space. Over 5,200 metric tons of cosmic dust arrives on the surface of the Earth annually [1]. The polymer has been extensively characterized from in-fall sources such as carbonaceous meteorites [2-7] and in fossil and present-day stromatolites [8]. The polymer may already have been detected by astronomers at wavelengths ranging from UV and visible to the infrared [9] in molecular clouds and protoplanetary discs. Sea foam is relatively stable and its collection soon after formation, at a shoreline, was pursued as a potential source of the in-fall polymer. Added support for such a polymer sample collection mechanism came from the existence of a surface micro-layer [10] that may be 50 microns thick carrying accumulated organic molecules and serving as a floating interface between the ocean and the troposphere. Sea foam's appearance on a remote Rhode Island shoreline was noted by the authors to mostly be confined to late autumn into early winter, and this could be related to the axial tilt of the Earth [11]. A physical resemblance between sea foam structure and hemoglycin vesicles was an additional reason to analyze this foam source. The method whereby hemoglycin has been extracted from meteorite and stromatolite sources via a tested solvation system [2], in which hemoglycin polymers accumulate at a solvent inter-phase, was applied here to its detection in sea foam.

Sea foam is referenced in a poem by Hesiod 730-700BC and therefore known to exist by the Greeks even if they gave it a mythological context: Θεογονία, Theogonía [12]. A thorough scientific report on sea foam goes back to Plateau in 1873 [13] who described the statics of liquids subjected to molecular forces only. The topological rules outlined by Plateau on foam were covered by Hill and Eastoe in 2017 in a review of the stabilization and destabilization of aqueous foams [14]. The Hill review covers both biological sea foam and man-made foams. The chemical make-up of sea foam from a wide variety of sites in Europe was studied by Mecozzi and Pietroletti [15], who applied UV-visible absorption, FTIR and FTNIR spectroscopy to build an extensive data base. FTIR data was particularly revealing as it allowed identification of aliphatic chains, esters and fatty acids, silicates, aromatics, and of particular interest in the present work, the amide I and amide II bands of protein in the region of 1500 – 1700cm$^{-1}$. The high protein content of sea foam near kelp beds has been presented by Velimirov [16,17] and beyond protein its wider chemical diversity is reviewed by Schilling and Zessner [10].

We hypothesized that the underlying structure of sea foam that only appears in late autumn to early winter in the USA NorthEast could depend upon hemoglycin arriving at the ocean surface via a seasonally increased in-fall from space at that time, implying that time-dependent biological components could in many cases be secondary additions to the original structural material of the foam. Bacterial content of sea foam is secondary [18]. For a recent view of the organic matter in sea foam, NOAA (National Oceanic and Atmospheric Administration) has an update from June 2024 [19].

**RESULTS**
21 samples of sea foam were collected from two Rhode Island (RI) beaches between September 23rd 2023 and February 27th 2024 (see Methods for the GPS locations and methods Table 4 for a sample list). An outlying sample, sample21, was collected in mid-August 2024 after a series of RI storms from the remnant of Hurricane Debby (Table 4) and this coincided with the yearly Perseid in-fall [20]. The typical faceted appearance of the individual foam cells, with 120 degree angles, is shown for sample21 (Fig.1). All samples in the 2-phase Folch extraction (Fig. 2 and Methods) produced a typical physical form consistent with the presence of the space polymer hemoglcyin, namely stable vesicles [3] that formed at the interphase layer (Fig. 3). Matrix-assisted laser desorption of ions (MALDI) mass spectrometry was performed on 4 samples with sample 9 giving the strongest molecular signals from a dense interphase layer (Fig. 4). The presence of high salt from the ocean hampered MALDI analysis in many samples because metal ions can cause the matrix compounds to form complexes [21], reducing matrix availability for the protonation of sample molecules. MALDI mass spectrometry is however the best method to obtain information on unknown molecular samples because the separation of molecules and molecule fragments is a single step process avoiding loss of material and the introduction of artifacts associated with other multiple-step mass spectrometry methods. Extensive foam (sample 17) on a shoreline is shown in Figure 5 where it extends far out to sea. Due to massive turbulence this sample was a yellow/brown color from sand and plant debris in the foam.

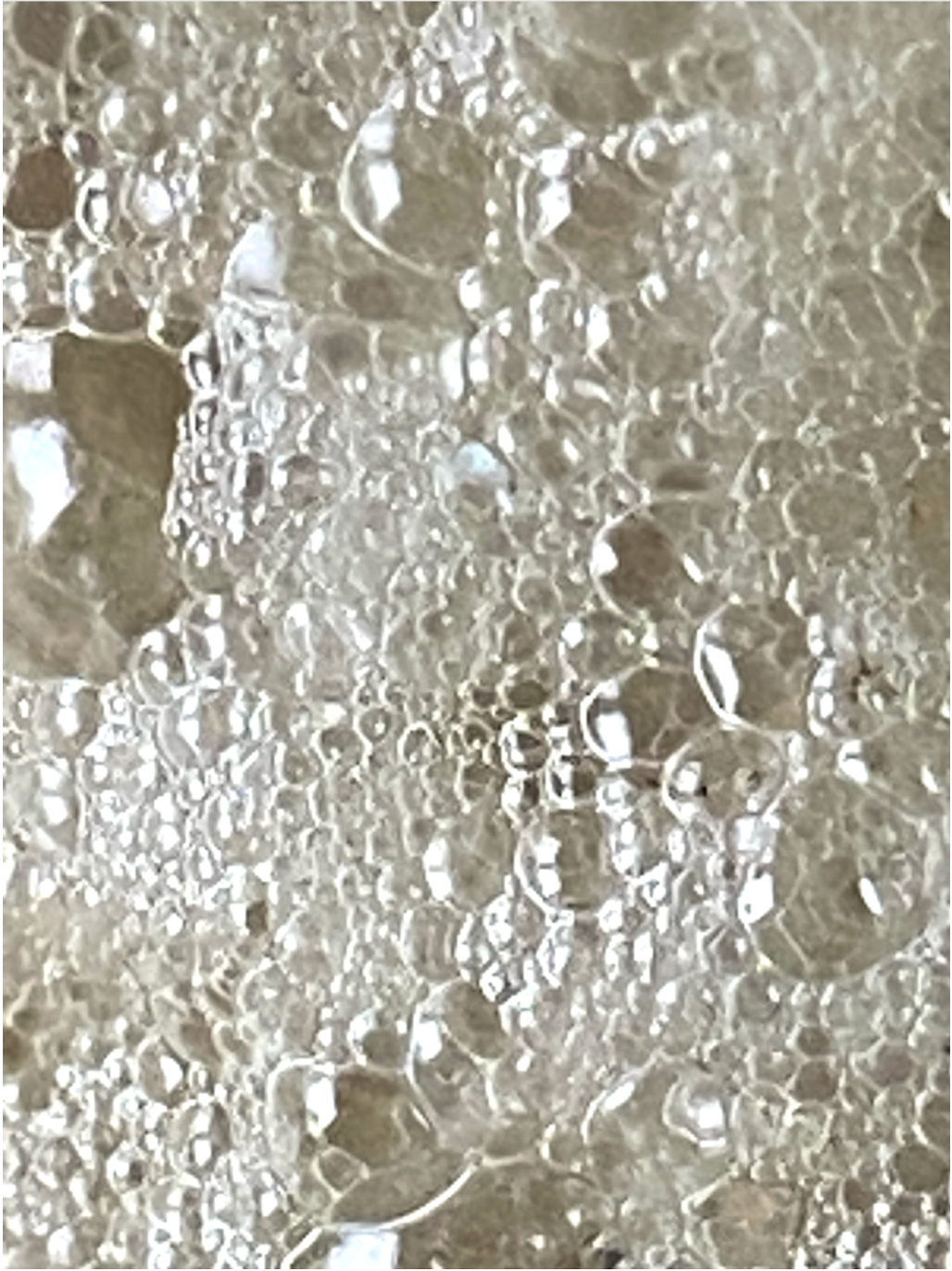

**FIGURE 1.** Sea foam sample 21, showing its faceted structure. Collected at the time point of the 2024 Perseid in-fall [20] and after a series of RI storms from the remnant of Hurricane Debby.

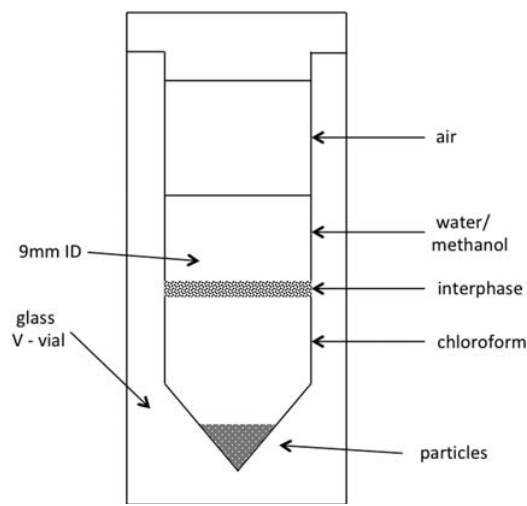

**FIGURE 2.** Folch extraction glass V-vial. 3D structures are at the interphase between water/methanol and chloroform. Particles sink in chloroform. From Julie E. M. McGeoch and Malcolm W. McGeoch. (2021) Structural Organization of Space Polymers. Physics of Fluids 33, 6, June 29th https://aip.scitation.org/doi/10.1063/5.0054860.

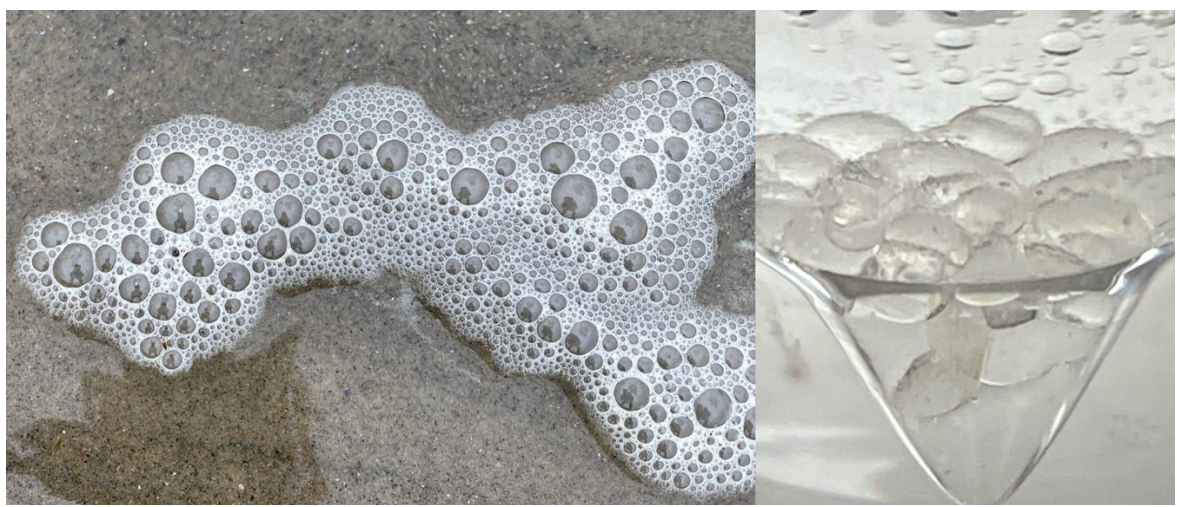

**FIGURE 3.** Clean floating sea foam (Sample 3) was collected directly from the shoreline of the ocean in RI on 23rd December 2023 (left). The wind was from the West. The foam was scooped in a glass bottle, transported to the laboratory, aspirated directly into chloroform in a V-vial and the methanol and water added to give a Folch 2-phase system of chloroform:methanol:water (3.3:2:1). On vortexing, clear vesicles formed at the Folch interphase (right). The vesicles were stable at room temperature for 1-2 weeks. The intact vesicles were analyzed by MALDI mass spectrometry producing from this sample only the 1657-1663 m/z sequence.

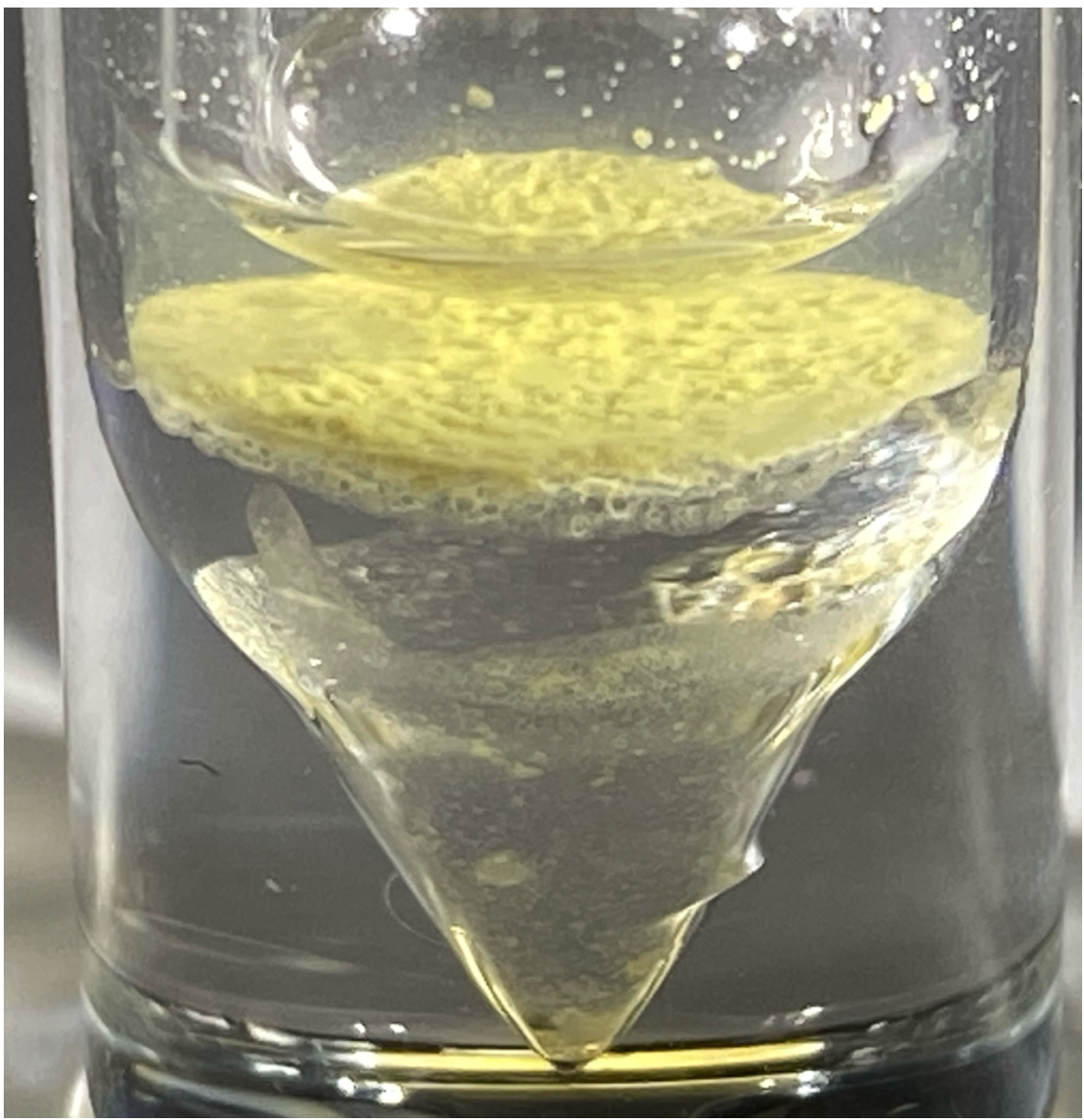

**FIGURE 4**: Folch extraction of Sea foam (Sample 9). Glass V-vial with yellow interphase Folch layer, chloroform below, water-methanol above, air at top.

Two controls were performed. At the exact point of the shoreline of sample 2 collection wet sand was collected 5 days later and subjected to Folch extraction but did not produce vesicles at the interphase. The 2$^{nd}$ control was sample 15 which was divided into two parts and one half heated in a crucible to 500C for 10 minutes. Both parts were Folch extracted with vesicles appearing at the interphase only in the unheated sample. A cylindrical yellow object adhered to the vial wall in sample 16 which was analyzed by MALDI however hemoglycin was not present in sample 16 but only matrix peaks were seen. The yellow object was considered to be plant contamination based on the fact it had a regular repeating structure within its length. Samples 12-15 had some sand present – all produced vesicles at the interphase plus some in the chloroform phase. Because hemoglycin tends to accumulate matter, fine sand particles made the vesicles too heavy to reside at the interphase layer.

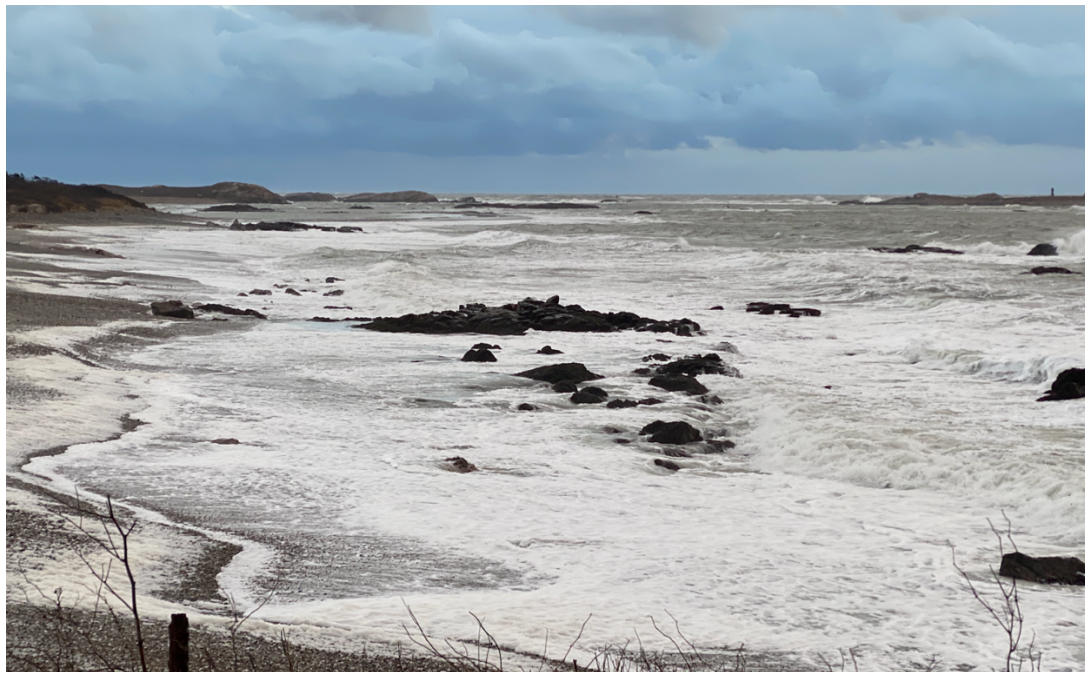

**FIGURE 5. Sea foam, sample 17, at Lloyds Beach location January 2024**. This sea foam stretched more than 100m out to sea. This location on this day produced foam that was not clean and contained visible sand and plant matter from the turbulent mixing of the sea by the wind and waves.

**Mass Spectrometry**
Our principal method of molecular analysis is MALDI mass spectrometry, through which intact hemoglycin, its adducts, and its typical fragmentation patterns have been determined in prior work [2,3,8]. In sample SF9, the analysis with highest signal-to-noise (S/N) ratio, we followed two principal fragmentation patterns that originated respectively from entities at m/z 1200 and m/z 1662. The m/z 1200 peak was the first of a series (Figure 6) rising in mass increments of 16, that represented successive degrees of oxidation of glycine residues to hydroxyglycine [3,6]. In work reporting the chiral 480nm absorption of hemoglycin [6] it was determined that a hydroxyglycine residue had to have its C-terminus adjacent to an iron atom in order to create the observed 480nm absorption. It was therefore assumed that any hydroxylation was present only in the four locations adjacent to terminal iron atoms, as shown for example in Figure 7. We discuss the 1200 and 1662 m/z entities in turn.
**The "1200" m/z series.**
In MALDI mass spectrometry there is typically seen a group of primary molecules that are extracted intact from the laser-heated matrix, providing an upper bound to the observed m/z values, and this group may be accompanied by lower m/z fragment peaks that help confirm its identity. The matrices, prepared as described in Methods, were CHCA (α-

cyano-4-hydroxycinnamic acid) and SA (sinapinic acid) and each matrix was used in duplicate, e.g. runs labelled SA-1 and SA-2.

With sample SF9, runs CHCA-1, CHCA-2, SA-1 and SA-2 there was a series of peaks beginning at m/z 1200 (Fig. 6) that provided an excellent case study of the MALDI fragmentation of hemoglycin. This series were seen with both matrices, CHCA (mass 189) and SA (mass 224), ruling out the possibility that they could relate to matrix clusters. The amplitude of the series decreased steadily as m/z rose through 1200, 1216, 1232, 1248 and 1264 suggesting a molecular type that could carry increasing levels of oxygenation. In our prior work with hemoglycin [3,8] we had determined that typically four glycine residues out of 22 were oxidized at the alpha carbon to hydroxyglycine, which has residue mass 73. The "1200" oxygenation series was suggestive of this, and a further aspect of the traces, their (-1) and (-2) isotopologue levels indicated that Fe was present at a level of two or more Fe atoms. The signal-to-noise (S/N) level of the 1200 series itself was poor, and not adequate to determine Fe content by the method described [3]. However, many of the fragments originating from 1200 m/z had higher S/N and were analyzed below to obtain Fe content. A trial molecule was written down in Fig. 7 based upon hemoglycin although having 18 residues as opposed to 22 in the usual antiparallel strands closed out by Fe atoms [3,6]. The context of this length variation will be discussed below.

**Table 1. Within each member of the 1200 m/z progression the sum of peak counts in a complex comprising mono-isotopic peak and satellites is similarly distributed with either CHCA or SA as the matrix. The principal fragments in the m/z 600 range depend to a degree on the choice of matrix.**

| m/z | Molecule | CHCA 1 | CHCA 2 | SA 1 | SA 2 |
|---|---|---|---|---|---|
| 1200 | $18Gly(FeO_2)_2 - 2H$ | 17,700 | 2,900 | 4,700 | 3,300 |
| 1216 | $17GlyGly_{OH}(FeO_2)_2 - 2H$ | 14,500 | 2,300 | 2,600 | 2,400 |
| 1232 | $16Gly2Gly_{OH}(FeO_2)_2 - 2H$ | 9,800 | 1,000 | 1,600 | 1,300 |
| 1248 | $15Gly3Gly_{OH}(FeO_2)_2 - 2H$ | 4,800 | 0 | 900 | 0 |
| 1264 | $14Gly4Gly_{OH}(FeO_2)_2 - 2H$ | 1,400 | 0 | 0 | 0 |
|  | Fragment |  |  |  |  |
| 600 | $9Gly(FeO_2) - H$ | 0 | 0 | 5,500 | 3,600 |
| 619 | $9Gly(FeO_2) + OH + H$ | 35,100 | 13,300 | 4,000 | 3,000 |
| 635 | $8GlyGly_{OH}(FeO_2) + OH + H$ | 30,300 | 8,100 | 0 | 1,100 |
| 88 | $FeO_2$ | 74,000 | 197,000 | 5,200 | 3,400 |
| 529 | $8GlyGly_{OH}$ | 31,200 | 7,400 | 2,000 | 1,500 |
| 531 | $9Gly + OH + H$ | 52,000 | 16,400 | 4,000 | 3,400 |

Analysis of the data in all four of the mass spectra relating to this sample (SF9) produced the consistent set of fragments listed in Table 1. These pointed to a break-up mode into two equal fragments of m/z in the region of 600, via separation along a line of hydrogen bonds between the antiparallel beta strands, plus breakage of the two remaining peptide-Fe bonds, shown by the dashed red line in Fig. 7. Subsequent fragmentation occurred via the vertical red dashed line in that Figure. The expected fragments are all observed, including fragments containing hydroxyglycine and the $FeO_2$ fragment that was identified by the exact $^{54}Fe$ (-2) isotope peak intensity at m/z 86. The "600" m/z series fragments of Table 1 via isotope spectrum analysis, Section S1, contain a single Fe atom, confirming the disposition of Fe atoms, one at each end of the 1200 species. In relation to the primary mode of splitting of hemoglycin, the same hydrogen bond un-zipping mode separating the anti-parallel polyglycine strands was found to be dominant in the splitting of the 22-residue core unit into two fragments at m/z 730 [3].

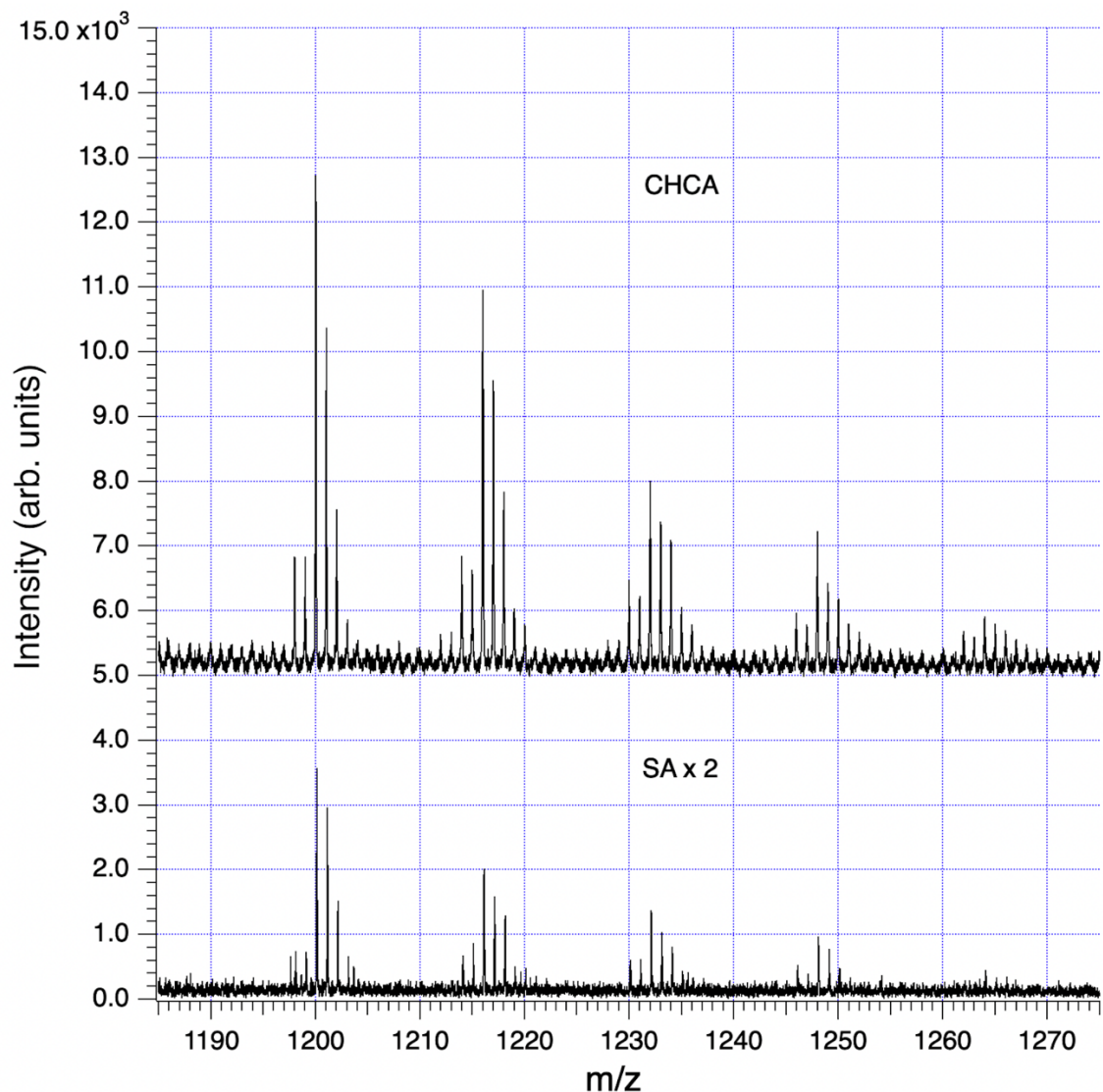

**FIGURE 6**. Series of hydroxylations in an 18-residue sea foam analog of the 22-residue hemoglycin core molecule, data in Table 1. The upper (CHCA) trace has been vertically offset for clarity. The lower (SA) trace is shown x2 for easier comparison.

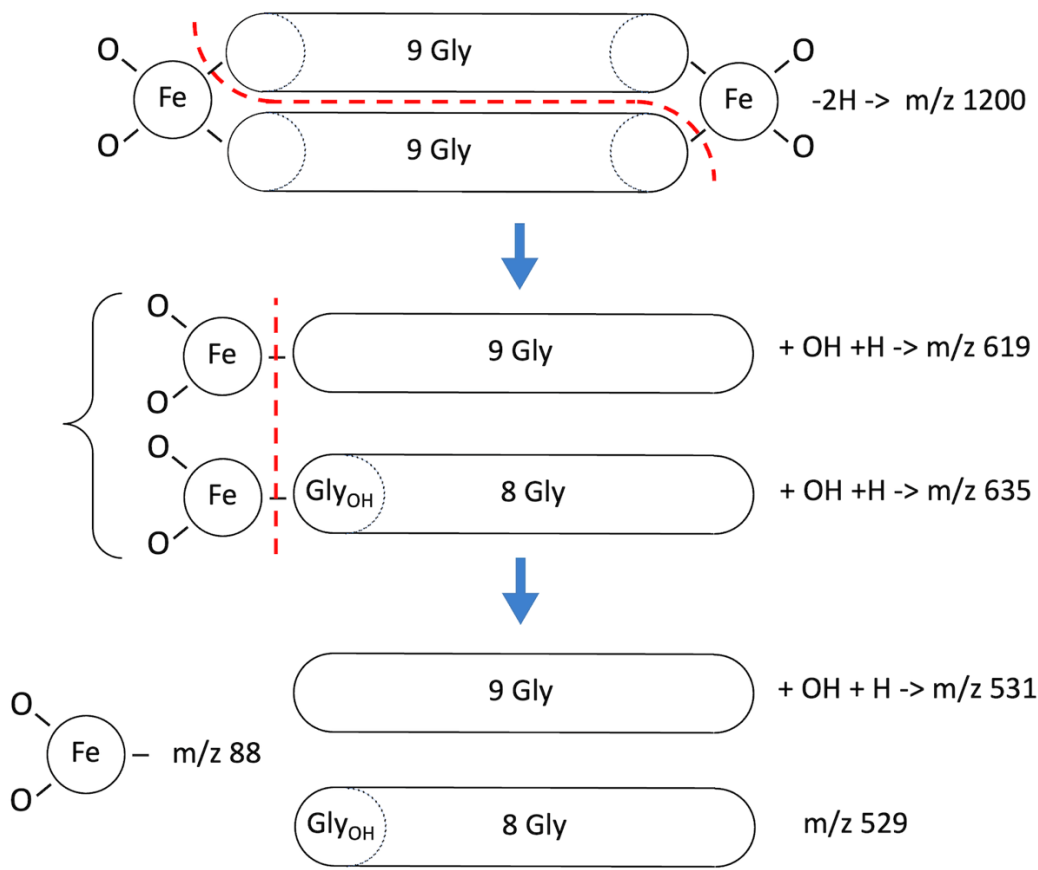

**FIGURE 7**. Fragmentation modes of the 1200 m/z series per Table 1. The single Gly$_{OH}$ residue in m/z 635 can be situated either adjacent to Fe (shown above), or distally.

**The "1662" m/z series**

The 1494Da "core" molecule of hemoglycin [3,6,8] can be observed directly in mass spectrometry from purified samples [8] but in the campaign that first revealed its existence [3] the two dominant species beyond the core molecule contained one additional FeO unit (to yield m/z 1567) or two FeO units (to yield m/z 1639). It was proposed in [5] that a three-dimensional lattice of the diamond 2H structure could have formed, and this structure was further confirmed in X-ray diffraction [8]. This structure has tetrahedral symmetry at all vertices, leading to the belief [5] that Si atoms with tetrahedral bond symmetry occupied these locations. With Si vertex atoms and additional oxygen atoms the (un-hydroxylated) 1638Da molecule [3] has augmented mass 1662Da as shown in Fig. 8 and Table 2.

**Table 2. Fragmentation data for the 1638Da hemoglycin primary molecule augmented to 1662Da for cases with or without hydroxyglycine (Gly$_{OH}$), illustrated in Figures 8 and 9. (*) last row, m/z 206 for CHCA and m/z 204 for SA.**

| m/z | Molecule | CHCA_1 | CHCA_2 | SA_1 | SA_2 |
|---|---|---|---|---|---|
| 1662 | 22Gly(Fe$_2$SiO$_4$)$_2$ | 0 | 0 | 0 | 0 |
|  | Fragment | counts |  |  |  |
| 489 | 5GlyFe$_2$SiO$_4$ | 254,000 | 74,900 | 17,400 | 6,000 |
| 505 | 4GlyGly$_{OH}$Fe$_2$SiO$_4$ | 280,000 | 60,500 | 9,400 | 5,000 |
| 360 | 6Gly + OH + H | 198,500 | 42,800 | 42,700 | 21,000 |
| 376 | 5GlyGly$_{OH}$ + OH + H | 161,300 | 24,200 | 53,500 | 29,800 |
| 190 | 3Gly + OH + 2H | 51,500 | 130,000 | 20,600 | 8,600 |
| 206 * | 2GlyGly$_{OH}$ + OH + 2H | 30,000 | 63,000 | 22,000 | 13,600 |

The principal fragments of the inferred m/z 1662 species, with either zero or one hydroxylation, are listed in Table 2. The signals were much lower with matrix SA and generally not at sufficient S/N to obtain definite Fe readings as found for example with CHCA in the 489 and 505 m/z species listed in Table 2.

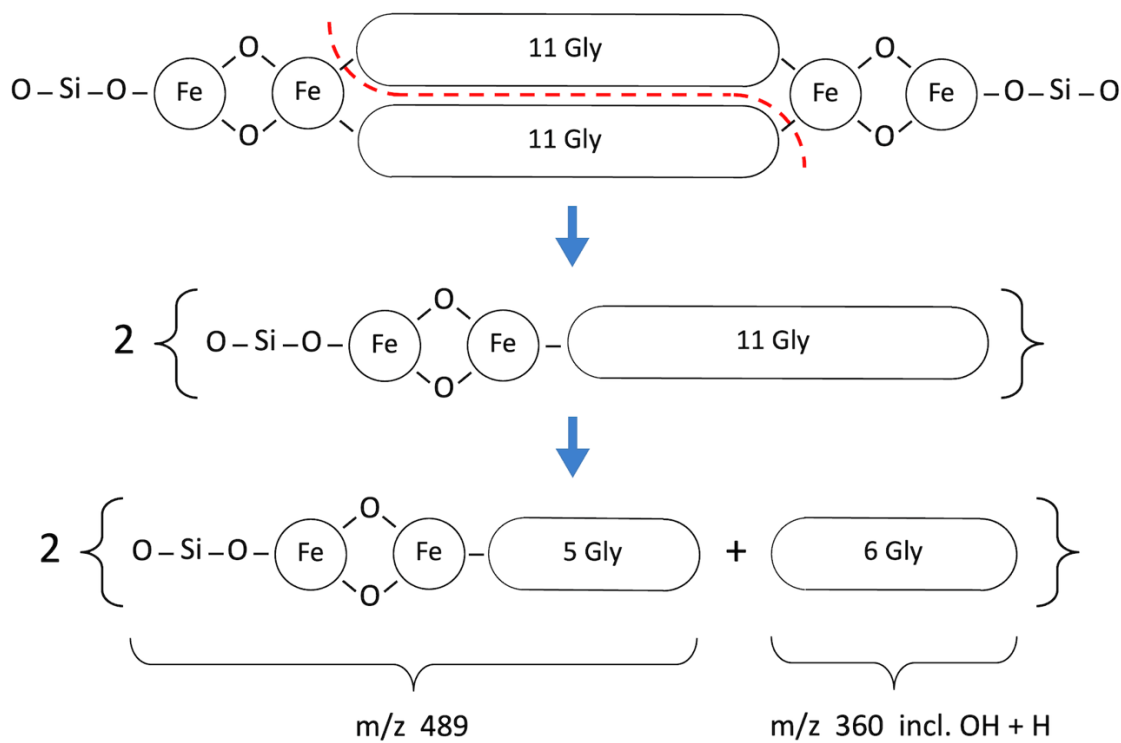

**FIGURE 8.** Fragmentation pattern for the augmented 1662Da hemoglycin lattice molecule in the case of no Gly$_{OH}$ residues. The primary break is symmetrical (red dashed line).

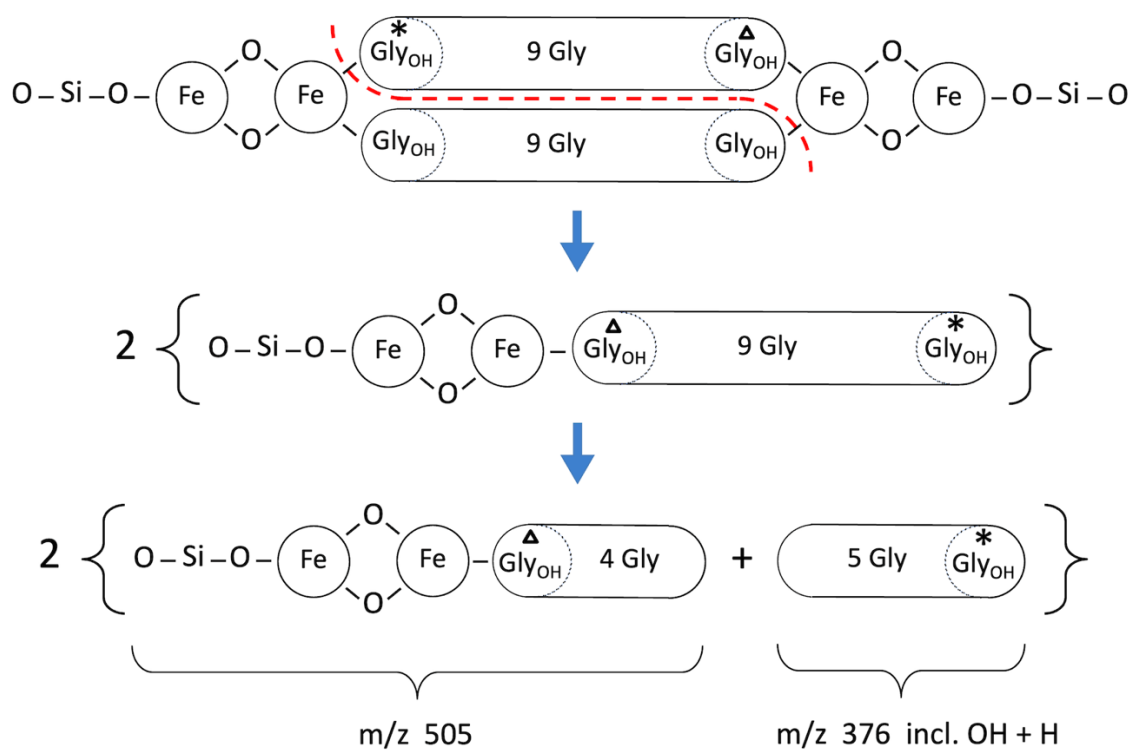

**FIGURE 9.** Fragmentation pattern for the augmented 1662Da hemoglycin lattice molecule in the case of labelled Gly$_{OH}$ residues that can be followed via identifying symbols. The primary break is a symmetrical splitting of the molecule (red dashed line).

In Figures 8 and 9 the primary break is again symmetrical (red dashed line), however different products are produced according to the number and disposition of Gly$_{OH}$ residues that can be followed via the identifying symbols. As important as finding the identity of a given fragment is verifying the absence of certain related fragments. For example, in determining the structure of m/z 489 and its single hydroxylation m/z 505, it was seen that m/z 521 was absent, indicating that two hydroxylations were not present at a significant level in any one beta strand of this primary molecule. Figure 10 presents a sample of the high S/N mass spectrometry traces that allowed Fe determinations via isotopologue analysis for the fragments in Figures 8 and 9.

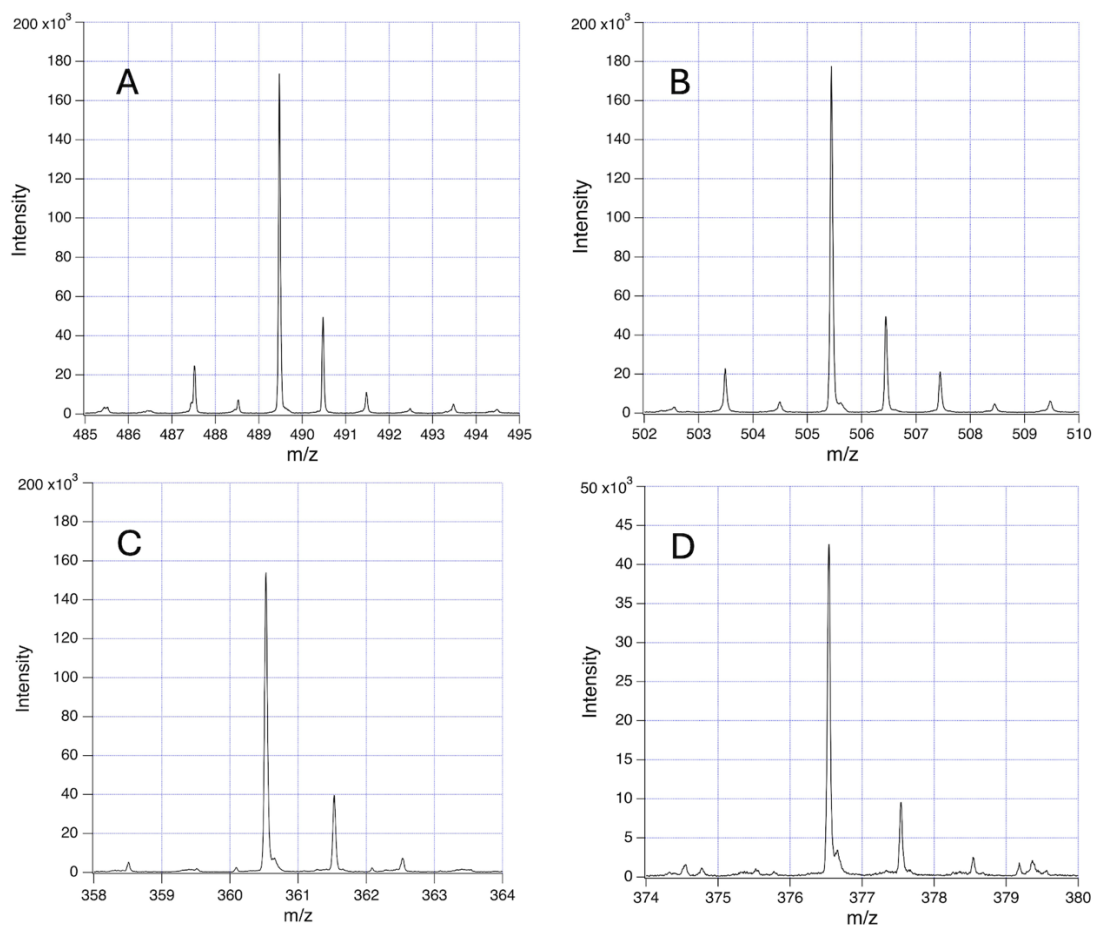

**FIGURE 10.** Illustrations of the isotopologue distributions for the fragments in Figures 8 and 9: A) 489 m/z; B) 505 m/z; C) 360 m/z , all with CHCA, and D) 376 m/z with SA.

**Observation of a "1662" m/z peak complex.**

Having traced in sample SF9 the fragment spectrum back to a 1662Da entity a corresponding peak complex was found in sample SF3, although only in the SA matrix cases. It stood alone as the highest peak complex for m/z exceeding 1190. This complex, shown in Figure 11, is at poor S/N but appears to have 3 or 4 Fe atoms, according to the relative height of the 1655/1656 isotopologues compared to the first large "mono-isotopic" peak at 1657 m/z. According to the fragment reconstruction in Figure 8 and Table 2 there was predicted a species $Gly_{22}Fe_4Si_2O_8$ (m/z = 1662). It appears that in MALDI up to 5 protons can be missing from this complex molecule, the charge being compensated by $Fe^+$ states. Apart from its low S/N, it appears that a variable number of proton adducts could be present, rendering isotope analysis impossible in this case either for Fe content or global $^2H$.

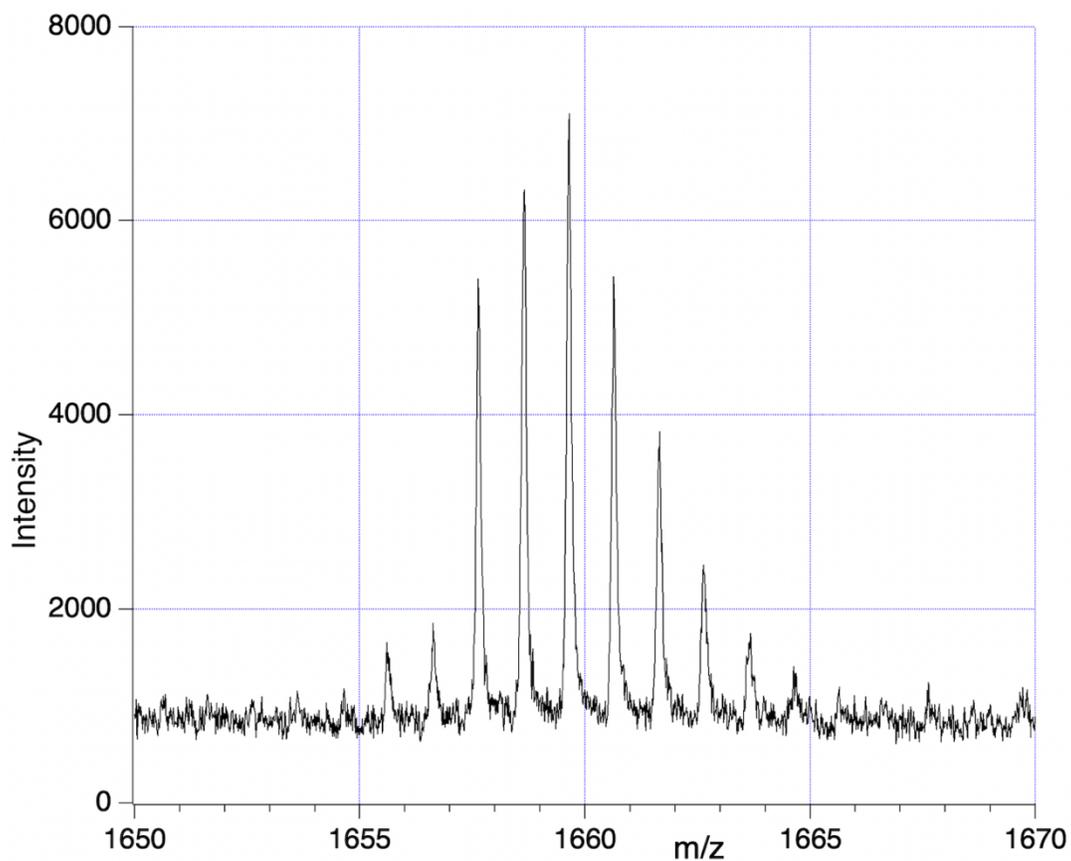

**FIGURE 11.** A complex molecular set observed in sample SF3 at approx. 1660 m/z.

**Discussion of MALDI fragmentation results**

The Fe content of fragments could be found by isotope analysis (S1) for all of the species listed in Tables 1 and 2, with the exception of the 1200 series for which S/N was inadequate. Two consistent fragmentation sequences (Figures 7, and 8+9) could be constructed that preserved original Fe configurations such as $FeO_2$ and $Fe_2SiO_4$ and also followed the previously observed "un-zipping" fragmentation mechanism of the 1494Da hemoglycin "core" molecule into m/z 730 moieties [3]. The "1200 m/z" fragment sequence originated in a reduced version of the core molecule with 9 glycine residues per chain as opposed to 11. Quantum chemical modeling [6] has shown that odd numbers of residues in the chain are more stable than even, due to the need to present converging free bonds toward the terminal Fe atoms. Only 11- residue chains have been observed in our prior work. The core chains in the present work are 9-residue (for 1200m/z) and 11- residue (for 1662m/z). The second fragment sequence derives from a 1662Da core entity that was a silicon adduct of the 1638Da molecule described in [3], itself based upon the "core" 22 residue hemoglycin core molecule but carrying four Fe atoms as opposed to two. The intensities of m/z peaks were very approximately five times smaller for the 1200 m/z set compared to the 1662 m/z set.

In each fragmentation sequence there were (+16) mass increments that are interpreted as varying degrees of hydroxylation of glycine. This was strikingly evident in the "1200" series (Fig. 6) where up to four hydroxylations were observed (Table 1). Prior mass spectrometry [3] of meteoritic samples had shown that four hydroxyglycine residues was the standard, and the observation of a chiral 480nm absorption [6] pinned down the position of at least one hydroxyglycine residue as being adjacent to the terminal Fe atom. Possibly because the sea foam chemical environment is "active", in water solution and exposed to visible/ultraviolet light, the proposed cycling mechanism associated with water-splitting [8] is being observed. In this mechanism each glycine residue adjacent to Fe cycles between plain glycine and hydroxyglycine, which could give a reduced average "occupancy" compatible with the present observations. At face value, for the "1200" series, the average number of hydroxylations is 1.0 in the CHCA spectrum, and 0.9 in the SA spectrum of Fig. 6. The rough equality of the 489 m/z and 505 m/z fragment counts in Table 2 would suggest an average of 0.5 hydroxylations across the two main fragments in that sequence, summing to 1 hydroxylation in the complete 1662 species that is the whole molecule counterpart of the 1200 species, in qualitative agreement.

**Isotope analysis in MALDI**
In the present paper, to give the isotopic analysis general utility we perform a "global" analysis as if $^2H$ were the only enriched isotope (Method in S1). If information subsequently becomes available either via a specific $^2H$ or $^{15}N$ measurement on a completely pure molecular sample then, under certain assumptions, the complementary enrichment of hydrogen or nitrogen can be estimated.

The enrichment of $^2H$ relative to $H$ is defined by the "per mil" $\delta$ ($^0/_{00}$) measure:

$$\delta^2 H \ (^0/_{00}) = \left( \frac{(^2H/H)_{SAMPLE}}{(^2H/H)_{VIENNA}} - 1 \right) \times 1000 \ \text{in which} \ (^2H/H)_{VIENNA} = 155.76 \pm 0.05 \times 10^{-6}$$

(IAEA, Vienna, 1995).

Controls for this isotope analysis were obtained from known matrix cluster peaks [21], giving a baseline of $700 \pm 1,700$ ($^0/_{00}$) enrichment for these terrestrial compounds (S1).

**TABLE 3. Isotope measurements on main fragment species. "Global" enrichment represented by $^2H$ per mil ($^0/_{00}$) enrichment from fitted isotopologue intensities.**

| run | m/z | Formula | Enrichment ($^0/_{00}$) |
|---|---|---|---|
| SF9 CHCA-2 | 360 | 6Gly + OH + H | 20,000 |
| SF9 (SA-1 + SA-2) | 360 | 6Gly + OH + H | 16,500 |
| SF9 CHCA-1 | 376 | 5GlyGly$_{OH}$ + OH + H | 22,500 |
| SF9 SA-1 | 376 | 5GlyGly$_{OH}$ + OH + H | 20,000 |
| SF9 CHCA-1 | 489 | 5GlyFe$_2$SiO$_4$ | 20,000 |
| SF9 CHCA-1 | 505 | 4GlyGly$_{OH}$Fe$_2$SiO$_4$ | 21,000 |
| SF9 CHCA-1 | 619 | 9GlyFeO$_2$ + OH + H | 15,000 |
| SF9 CHCA-2 | 619 | 9GlyFeO$_2$ + OH + H | 17,500 |
| SF9 CHCA-1 | 635 | 8GlyGly$_{OH}$FeO$_2$ + OH + H | 12,500 |
|  |  | n = 9    ave. = 18,300    σ = 3,000 | |

**DISCUSSION**

This research suggests that the underlying structure of sea foam involves the polymer hemoglycin. The polymer could first form in the Universe once its constituent elements existed 13 billion years ago [5,9]. The earliest references to sea foam ("aphros" = foam) occur in the Greek myth of the heavenly origin of Aphrodite [12] without the then support of mass spectrometers, accurate telescopes and X-ray diffraction from synchrotrons. That relatively clean sea foam collected from two Rhode Island beaches contains the space polymer hemoglycin is deduced from the presence of vesicles after Folch extraction and from MALDI mass spectrometry. Vesicles present at the Folch extraction "interphase" layer are a key visual indicator that hemoglycin is in a sample, be it from meteorites [3], stromatolites [8] or, in the present case, sea foam. The polymer of glycine and iron adopts a position in the two-phase Folch extraction interphase relative to its density and hydrophobic/hydrophilic structure. It is less dense than chloroform, residing at the top of the hydrophobic chloroform layer but remaining below the less dense hydrophilic methanol-water layer. Sand on the shoreline near to where foam landed did not contain hemoglycin and samples heated to 500C had the hemoglycin destroyed. The sea foam samples analyzed thus had properties that indicated they contained a molecule sensitive to heat and were not from a sand source at the shore-line. Space polymer in 100micron scale in-fall particles potentially remains at the ocean surface and together with surface layer surfactants is incorporated via turbulent wave action into a foam, extracts of which retain the elevated heavy isotope levels characteristic of meteoritic or cometary material.

Cosmic material reaches Earth's surface as meteorites and also, more steadily and in larger quantities, as cosmic dust. Dust incident at orbital height above Earth's atmosphere has been measured by space probe to arrive at a rate of $4 \pm 2 \times 10^4$ metric tons per annum [22] with a size distribution peaking at 200microns diameter. Particles entering the atmosphere are heated by ablation in some cases to melting point, producing spherules, and in other cases losing mass by evaporation (categorized as un-melted micro-meteorites, or uMM). Combined averaged-over-time cosmic dust deposition in clean snow at Dome C, Antarctica [1] amounts to 5,200 +1,500/-1,200 tonnes per annum when extrapolated to the whole globe, with size distribution peaking at about 100microns diameter. In 26 years of walks on the two RI beaches we have only noted sea foam in the Autumn and early winter months and indeed the samples of the present work were gathered between 24[th] Oct 2023 and 27[th] Feb. 2024. (Table 4, sea foam samples). An exception to this pattern was the current year of 2024 where several spring to summer storms occurred. Two sea foam samples collected in mid-August 2024 after very turbulent seas due to the remnants of Hurricane Debby, produced typical vesicles after Folch extraction (Table 4). In search of an explanation we note that, because of the fixed tilt of Earth's axis [11], the Northern hemisphere is approaching the "zodiacal cloud" of meteoroids at one equinox, and the Southern hemisphere at the complementary equinox. Many other variables may be at work, such as more storms to agitate the sea surface in winter, or different surface microlayer surfactant content when the sea is colder. Also, meteor shower debris could contribute at other times.

As to the composition of in-falling dust, the picture can be complex, involving five major categories according to radar orbital analysis, with the principal category, amounting to 68% of the total, being the helion-antihelion fluxes [23]. Containing cometary material, often from Jupiter family comets, these pass Earth en-route to perihelion close to the sun, and re-pass on the outward-bound leg at slightly reduced intensity. This group has relatively short collisional lifetime and must be constantly replenished by new cometary material on timescales of the order of $10^4$ years in order to remain the dominant apparent source of meteoroids [23]. The properties of ultra-carbonaceous Antarctic micro-meteorites (UCAMMs) [24] lend support to a cometary origin, for at least a fraction of the in-fall, in respect to their high N/C ratio.

Fine-grained uMM particles and UCAMMs have the highest content of "unaltered" carbon, within surviving organic molecules, and this may amount to 50 tonnes per annum when extrapolated to the globe [1], although this number is uncertain. The composition of the UCAMM organic material [24] differs from carbonaceous meteorites in having a very high N/C ratio. Also, it has a strong infrared absorption in the region of 1600-1700 $cm^{-1}$, a range corresponding to the amide I absorption of anti-parallel beta sheets of polymer amide [9]. In the [24] data there is also a prominent N-H vibrational mode absorption in the region of 3,300 $cm^{-1}$ that has been observed in FTIR of a meteoritic sample containing hemoglycin [8,9]. While not providing conclusive proof that meteoroid infall is responsible for hemoglycin in sea foam, these facts are supportive.

As to the fate of in-falling 100micron meteoroid particles of density about 2, it seems possible for many to be attracted into the sea surface layer of molecules known as the surface-microlayer (SML) [10] which can be 50 microns thick and contains a wide range of organics, most resulting from the decay of organisms, as well as protein contributions by vegetation such as kelp [16,17]. First contact of meteoroid dust with sea water will cause alkali metals to react, buoying the particle by gas release, and potentially causing organics to leach into the SML. Spherules, on average more massive and having smoothed surfaces, are more likely to sink to the sea floor. Infrared absorption of sea foam samples [15] yields dominant amide I bands in the 1620 – 1700 $cm^{-1}$ region, consistent with the presence of hemoglycin [9] or other proteins with an anti-parallel beta sheet structure. Present-day stromatolite ooids contain hemoglycin as evidenced by x-ray induced fluorescence, x-ray scattering patterns and infrared absorption [8], thus confirming the availability of in-fall hemoglycin on a current basis in shallow, clear saline waters.

Isotope enrichment is observed in hemoglycin molecules identified by mass spectrometry [3,8] and the same isotope analysis is here applied to the identified polymer amide molecules in Folch extracts of sea foam (details in S1). Two heavy isotopes are substantially raised in cometary and primitive solar system material, namely $^{15}N$ and $^{2}H$ [4]. In polymer amide from the Acfer 086 meteorite the $^{15}N$ enrichment was 1,015 ($^0/_{00}$) via secondary ion mass spectrometry (SIMS) [4], in the same category as cometary $^{15}N$ enrichments [25]. $^{2}H$ enrichments of identified hemoglycin molecules are substantial, in the range 25,000 to 50,000 ($^0/_{00}$), the latter number obtained from the Orgueil meteorite [8]. If, as we believe, the sea foam hemoglycin is in-fall material, its original peptide backbone intact, then it still carries its original $^{15}N$ enrichment trapped within in its repeating -CCN- backbone. On the other hand, the $^{2}H$ atoms of hemoglycin slowly exchange with $^{1}H$ of sea water. A distinction is made between the relatively rapid exchange of hydrogen in H-N bonds, and hydrogen in H-C bonds involving the alpha carbon of the peptide backbone.

Overall, the rate of D-H exchange depends upon solution pH, and peptide conformation. For example, [26] found a variable percentage of hard-to-exchange amide hydrogens (HEAHs) that increased to as much as 60% in insulin, attributed to a high content of strongly hydrogen-bonded secondary structures. Polyglutamic acid exchange at pD 3.5 did not reach completion in 168 hours, whereas N-deuterated proteins exchange relatively rapidly [27]. In a simulation of the alkaline environment for meteoritic extraction [28] measured a D-H exchange rate in dicarboxylic acid C-H bonds of $(1.3 – 8.4) \times 10^{-4}$ /hr in solution at 100C. In summary, intact hemoglycin molecules are expected to retain their backbone $^{15}N$ content, but to gradually lose their D content while afloat in the sea surface layer. Our data provides a "global" enrichment of $18,300 \pm 3,000$ ($^0/_{00}$) that includes $^{15}N$ and $^2H$, within which there is likely to be the cometary $^{15}N$ of $1,040 \pm 90$ ($^0/_{00}$) [25]. The hemoglycin molecule with H/N = 3 has a $^2H$ decrease of 7.73 x (increase of $^{15}N$) ($^0/_{00}$) (S1), so to assume cometary $^{15}N$ in the backbone leads to an estimated decrease of 8,040 ($^0/_{00}$) to global enrichment, giving $10,300 \pm 4,000$ ($^0/_{00}$) as a conditional estimate for $^2H$ residual enrichment in sea foam hemoglycin. The removal of most of the expected 25,000 – 50,000 ($^0/_{00}$) $^2H$ enrichment for cometary hemoglycin (via [3], and Orgueil data [8]), thus implies that most of the deuteration in original cosmic dust may have been exchanged for seawater $^1H$ by the time samples were collected.

Hemoglycin is present in sea foam, but does it contribute to the foam structure, or is it merely present as one of many organic components? The ready formation of vesicles by hemoglycin, mentioned above, could allow it to contribute to the structure of the very thin walls throughout sea foam. In prior work [3] and an unpublished observation of x-ray scattering (S2) we have found that hemoglycin rods can form a planar hexagonal lattice with side length of about 4nm, the length of a 1494Da core hemoglycin unit. Such an effectively two-dimensional lattice can, with small admixtures of pentagons or hexagons, cover arbitrary curved surfaces. In the S2 data, a crystal of hemoglycin derived from the Sutter's Mill meteorite [6] contained a region that scattered x-rays into a simple set of six directions, the pattern persisting over a large range of incident angles, leading to the identification of scattering by a lenticular form, or collapsed vesicle, with multiple layers of this hexagonal lattice covering its surface. It is therefore possible to imagine that hemoglycin would be able to reinforce surfactant layers, allowing them to be very stable even as water drained from them, creating the very light and rugged foam that we observe. Finally, in regard to the propensity for sea foam to detach and fly in the wind, we have identified a possible water-splitting reaction cycle in which the hydroxyglycine next to a terminal iron atom can in two stages [8] effect the overall reaction $2H_2O + 2h\nu \rightarrow H_2 + H_2O_2$, partially filling sea foam cavities with hydrogen.

**METHODS**
Collection of Sea Foam samples
Sea foam was collected from 2 locations:
1. Samples 1-21 except sample 17. Warrens Point Beach, Little Compton, RI 02837
GPS: **Latitude:** 41.481924 | **Longitude:** -71.144582.
2. Sample 17. Lloyd Beach, Little Compton RI 02837.
GPS: **Latitude:** 41.461357 | **Longitude:** -71.196024.

The time of collection was determined by charts from US Harbors [wwww.USH.com] on a daily basis from December 2023 to February 2024. Clean foam was obtained when a minimum low tide mark had been reached and the tide was incoming by 1-2 hours.

Sea foam is very stable and was collected in a wide necked glass container by scooping the foam from the sea directly into the container. It needs to be noted that once a small amount of foam is detached from a mass of foam floating on the sea or on the beach, it tends to become rapidly airborne, we believe because it contains a very low-density gas, probably hydrogen. Catching the foam in the container required positioning the container to use the wind direction to ensure the foam entered the container. It was then transported to the laboratory, chloroform added to dissolve, and poured into a glass V-vial. Methanol and some water were added, the whole V-vial gently vortexed to produce foam vesicles at the interphase of the chloroform:methanol:water (3.3:2:1) Folch extraction used previously for meteoritic samples [2] [Fig.2]. Figure 3 shows a small amount of sea foam that was very clean, apparently containing only salts and hemoglcyin. Figure 4 shows collection and handling of a sample with a more dense amount of hemoglycin plus other secondary chemicals. Figure 5 shows a very large amount of foam on the Lloyds beach location where the foam stretches hundreds of feet out to sea. This sample had many secondary substances present like sand and plant matter from the turbulent sea and represents a sample not analyzed because of contamination.

**MALDI Mass Spectrometry of Sea Foam Samples**

Intact stable vesicles from the Folch interphase layer were aspirated from the Sample SF3 V-vial (Fig.2) into duplicate Eppendorf tubes containing the CHCA ($\alpha$-cyano-4-hydroxycinnamic acid) or SA (sinapinic acid) matrix mixes described below. For Sample SF9 (Fig.4) 200μl of the yellow interphase layer were pipetted into duplicate Eppendorf tubes containing either CHA or SA matrix mix.

Mass spectrometry was performed on a Bruker Ultraflextreme MALDI-TOF/TOF instrument in the positive ion, reflectron mode. We used CHCA and SA matrices. Both were at 10 mg/mL in 50% acetonitrile in water, 0.1% trifluoroacetic acid in water. Our resolution was of the order of 10,000 and we looked in the range m/z = 0–5,000, finding most peaks from 20-2,000. A sample volume of 2μL was mixed with a matrix volume of 2μL, vortexed and left for one hour at room temperature. This one hour wait before pipetting 1μL quantities onto the MALDI plate is essential to partly solubilize hemoglycin in the matrix solvents.

**X-ray structural analysis (Section S2)**

On Diamond beamline 124 diffraction data were recorded from the Sutter's Mill SM2 crystal at 1.000A, 0.1$^O$ oscillation range, using a Pilatus3 S 6M detector.

**Table 4 Sea Foam samples.**

| Sea foam sample | Date of collection | Stable vesicles at Folch interphase | MALDI mass spectrometry |
|---|---|---|---|
| 1 | 23.10.24 | Yes | |
| 2 | 23.11.21 | Yes | |
| CONTROL sand collected near to sample 2 location | 23.11.26 | NO | |
| 3 | 23.12.2 | Yes | Yes on vesicles at interphase **Hemoglycin detected** |
| 4 | 23.12.10 | Yes | |
| 5 | 23.12.11 | Yes | |
| 6 | | Yes | |
| 7 | 23.12.19 | Yes | |
| 7 | 23.12.20 | YES | |
| 7 | 23.12.21 | YES | |
| 8 | 23.12.23 | SF introduced to just chloroform – vesicle bounding sea water above chloroform | Yes on dry crystal after chloroform evaporated after 22days. Hard crystal not solubilized by matrix mix |
| 9 | 23.12.23 | Yes | Yes on interphase samples **Hemoglycin detected** |
| 10 | 23.12.27 | Yes | |
| 11 | 24.1.3 | Yes | |
| 12 sand present | 24.1.6 | Yes & denser vesicles in chloroform phase | |
| 13 sand present | 24.1.7 | Yes & denser vesicles in chloroform phase | |
| 14 sand present | 24.1.7 | Yes & denser vesicles in chloroform phase | |
| 15 sand present Sample divided into 2: 1) Heated CONTROL 2) Unheated test Then both Folch extracted | 24.1.11 | No vesicles in sample heated in crucible for 10minutes 500C. Yes vesicles in test | |
| 16 | 24.1.11 24.1.30 | Yellow object on vial wall | Yes but only matrix peaks |
| 17 | 24.1.14 | Yes | |
| 18 | 24.1.31 | Yes | |
| 19 | 24.2.26 | Yes | |
| 20 | 24.2.27 | Yes | |
| 21 | 24.8.9 & 10 | Yes | Storm from the south coincided with the yearly Perseid in-fall. |

ACKNOWLEDGEMENTS

We thank Sunia Trauger, the senior director of the Harvard Center for Mass Spectrometry for assisting with the MALDI experiments and Smithsonian Astrophysical Observatory Center for Astrophysics, Harvard & Smithsonian, for funding the mass spectrometry. For the S section the authors thank Diamond Light Source UK for beamtime (proposal MX31420) and Robin Owen, Sofia Jaho and Sam Horrell for hemoglycin crystal SM2 X-ray data collection on beamline I24. The Sutter's Mill meteorite sample was kindly provided by Michael E. Zolensky of NASA.

**Funding:** This work was supported in part by the Center for Astrophysics, Harvard & Smithsonian for (J.E.M.Mc);

**Declaration of interests:** the authors report no conflict of interest.

**Data availability:** The data that support the findings of this study are available from the corresponding author upon reasonable request and at the Harvard Dataverse repository via URL: https://doi.org/10.7910/DVN/A00GMD,

**Author ORCID:** JEMMcGeoch, https://orcid.org/0000-0002-4319-9836. MWMcGeoch. https://orcid.org/0000-0002-3162-8541

# S Section: Sea Foam contains Hemoglycin from Cosmic Dust.

Julie E. M. McGeoch[1] and Malcolm W. McGeoch[2]

[1] High Energy Physics DIV, Smithsonian Astrophysical Observatory Center for Astrophysics | Harvard & Smithsonian, 60 Garden Str, MS 70, Cambridge MA 02138, USA
[2] PLEX Corporation, 275 Martine Str, Suite 100, Fall River, MA 02723, USA

## S1 Isotope analysis

**Molecular isotope analysis**

In our prior work with meteoritic extracts [1, 2], or space-derived materials such as the hemoglycin lattice within stromatolites [2], the isotopologues around a "mono-isotopic" mass spectral peak have intensities that relate to the isotopic composition of each of the atoms in the molecule. The significant heavy isotopic enrichment of extraterrestrial glycine polymers becomes plainly visible in the isotopologue distributions for m/z values as low as 400, while at higher m/z the (+1), (+2), etc. peaks often rise even above the (0) "mono-isotopic" peak. Provided the molecule's atomic composition is known from fragmentation (e.g. MS/MS) studies, and it is known which elements have enrichment, a measure of a specific isotopic enrichment can be derived via a straightforward simulation [1]. Discussion in [1, S section] points to the likelihood that $^{15}N$ and $^{2}H$ are the two principal highly enriched isotopes in polymer amide extra-terrestrial material. The enrichment of $^{15}N$ can be isolated via separate study of CN negative ions from the molecule using secondary ion mass spectrometry (SIMS). We have measured a $^{15}N$ enrichment of $1{,}015 \pm 280$ ($^0/_{00}$) in polymer amide within the Acfer 086 meteorite, in broad agreement with cometary estimates [3], and it is presumed that because N is an integral part of the –CCN- repeating polymer backbone, any intact polymer in-falling to Earth, such as found in the present study, will by force contain the original extra-terrestrial $^{15}N$ within its backbone. The variations in $^{13}C$ are expected to be relatively less consequential in this molecular type [1, 3].

In the present paper, to give the isotopic analysis general utility we perform a "global" analysis as if $^{2}H$ were the only enriched isotope. If information becomes available either via a specific $^{2}H$ or $^{15}N$ measurement on a completely pure molecular sample then, under certain assumptions, the complementary enrichment of hydrogen or nitrogen can be estimated.

The enrichment of $^{2}H$ relative to $H$ is defined by the "per mil" $\delta$ ($^0/_{00}$) measure:

$$\delta^2 H\ (^0/_{00}) = \left( \frac{(^2H/H)_{SAMPLE}}{(^2H/H)_{VIENNA}} - 1 \right) \times 1000 \quad \text{in which } (^2H/H)_{VIENNA} = 155.76 \pm 0.05\ \text{x}\ 10^{-6}$$

(IAEA, Vienna, 1995).

Without separate knowledge of $^{15}N$ enhancement in hemoglycin we fitted the prominent isotope enrichments as if $^{2}H$ was the only contributor. When $^{15}N$ values come available the quoted "global" $^{2}H$ enrichments may be modified as follows where $\Delta$ represents the degree of change:

$\Delta(\delta^2 H) = -7.73\Delta(\delta^{15} N)$. This relationship holds for polymers of glycine and hydroxyglycine that have $N_H/N_N = 3$. It may be scaled for general ratios of $N_H$ to $N_N$ via the relationship

$\Delta(\delta^2 H) = -(R_N N_N / R_H N_H)\Delta(\delta^{15} N)$ using the $R_N$ and $R_H$ values listed below.

The following isotope ratios (IAEA, Vienna, 1995) are taken as terrestrial standards:
VSMOW water $\qquad R_H = {}^2H/{}^1H = 155.76 \pm 0.05 \times 10^{-6}$
VSMOW water $\qquad R_O = {}^{18}O/{}^{16}O = 2,005.20 \pm 0.45 \times 10^{-6}$
V-PDB $\qquad R_C = {}^{13}C/{}^{12}C = 11,237.2 \times 10^{-6}$
Atmospheric Nitrogen $\qquad R_N = {}^{15}N/{}^{14}N = 3,612 \pm 7 \times 10^{-6}$

Figure S1.1 shows a typical fitted "global" isotope enrichment, in $^2H$ equivalent. The curve is for an assumed 21,000 mil (‰) $^2H$ enrichment with all other isotopes set at terrestrial (Vienna) level. The (-1) and (-2) components are at the level indicative of 2 Fe atoms in the m/z 505 entity. In the main text the possibility of a $^{15}N$ component to the global isotope enhancement is considered.

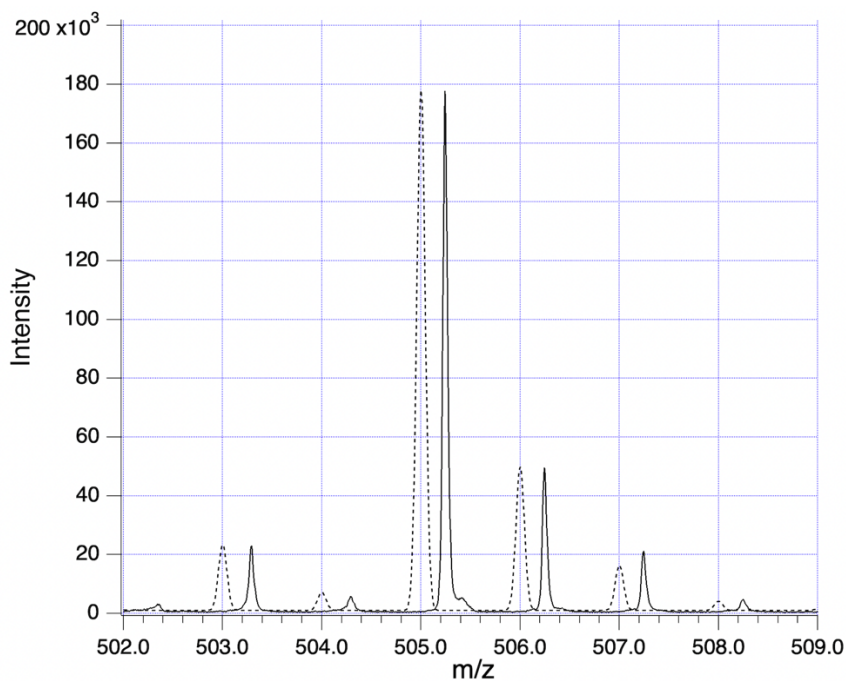

**Figure S1.1. Isotope match of the m/z 505 fragment in sample run SF9-CHCA1, showing 2Fe via $^{54}$Fe content at m/z 503. Data, solid line. Fitted curve, dashed line, normalized at peak and offset for clarity.**

Global isotope enrichments for the main fragment species are listed in Table S1.1. There is an average global "$^2H$" enrichment of 18,300 ± 3,000 (‰), which appears to indicate significant isotope enrichment in the hemoglycin component of sea foam (which is not to say that it would be measurable in sea foam as a whole, with its many additional organic components).

A convenient control for isotope levels in the m/z range of interest comes from MALDI matrix clusters that are well-characterized [4] and terrestrial to the accuracy required for the present work. We found high S/N matrix clusters in CHCA for three samples that had lower hemoglycin content and largely empty spectra apart from matrix peaks (SF3, SF8 and SF16). The isotope analysis was applied to these matrix clusters of known atomic contents with results shown in Table S1.2. Without any change to the Vienna set (above) the fits all indicated a small negative $\delta^2H$ except for one. Negative values were not included in the average because the enrichment becomes very nonlinear – for example, isotopically pure H would have $\delta^2H$ = -1,000 ($^0/_{00}$). Variation of $\delta^{13}C$ to -100 ($^0/_{00}$) is sufficient to fit most of the matrix cluster isotope patterns at otherwise terrestrial (Vienna) values. The positive global enhancements in sea foam hemoglycin (Table S1.1) are therefore statistically real, although their isotopic breakdown is not available from this approach.

**Table S1.1 Isotope measurements on main fragment species. "Global" enrichment represented by $^2H$ per mil ($^0/_{00}$) enrichment from fitted isotopologue intensities.**

| run | m/z | Formula | Enrichment ($^0/_{00}$) |
|---|---|---|---|
| SF9 CHCA-2 | 360 | 6Gly + OH + H | 20,000 |
| SF9 (SA-1 + SA-2) | 360 | 6Gly + OH + H | 16,500 |
| SF9 CHCA-1 | 376 | 5GlyGly$_{OH}$ + OH + H | 22,500 |
| SF9 SA-1 | 376 | 5GlyGly$_{OH}$ + OH + H | 20,000 |
| SF9 CHCA-1 | 489 | 5GlyFe$_2$SiO$_4$ | 20,000 |
| SF9 CHCA-1 | 505 | 4GlyGly$_{OH}$Fe$_2$SiO$_4$ | 21,000 |
| SF9 CHCA-1 | 619 | 9GlyFeO$_2$ + OH + H | 15,000 |
| SF9 CHCA-2 | 619 | 9GlyFeO$_2$ + OH + H | 17,500 |
| SF9 CHCA-1 | 635 | 8GlyGly$_{OH}$FeO$_2$ + OH + H | 12,500 |
|  |  | n = 9    ave. = 18,300 | σ = 3,000 |

**Table S1.2. Control isotope measurements on CHCA matrix species. "Global" enrichment represented by $^2H$ per mil ($^0/_{00}$) enrichment from fitted isotopologue intensities. M = CHCA = $C_{10}H_7NO_3$.**

| run | m/z | Formula | Enrichment ($^0/_{00}$) |
|---|---|---|---|
| SF16 CHCA-1 | 445 | M$_2$Na$_3$H$_{-2}$ | < 0 |
| SF16 CHCA-2 | 445 | M$_2$Na$_3$H$_{-2}$ | < 0 |
| SF16 CHCA-1 | 656 | M$_3$Na$_4$H$_{-3}$ | < 0 |
| SF8 CHCA-2 | 656 | M$_3$Na$_4$H$_{-3}$ | < 0 |
| SF3 CHCA-1 | 861 | M$_4$KNa$_3$H$_{-3}$ | < 0 |
| SF16 CHCA-2 | 861 | M$_4$KNa$_3$H$_{-3}$ | 5,000 |
| SF3 CHCA-1 | 877 | M$_4$K$_2$Na$_2$H$_{-3}$ | < 0 |
|  |  | n = 7    ave. = 700 | σ = 1,700 |

# S2. Meteoritic evidence for a hemoglycin hexagonal surface-covering mesh

## S2.1 Introduction

In this section we discuss a possible structural role for hemoglycin in the stabilization of sea foam bubbles. Its first chemical identification [1] was in CV3 class carbonaceous meteorites, particularly Acfer 086. A crystalline extract via Folch solvent from Sutter's Mill meteorite showed the characteristic square mesh structure of hemoglycin via its x-ray scattering [5]. However, in one crystal location, it displayed an extremely simple hexagonal X-ray pattern that we identified as being from a vesicle trapped within the sheet crystal. The vesicle appears to consist of a surface-covering hexagonal mesh composed of triskelia, or three-legged subunits, in which each "leg" is based upon a hemoglycin core molecule. The ability of hemoglycin to form curved surface sheets may contribute to the formation and stabilization of sea foam, hence the following summary of this unpublished X-ray observation.

Solvent extracts from meteorites frequently display floating small globules with a dark surface layer that remain stable for months [1], referred to here as vesicles (Fig. S2.1). Previously, X-ray data from hemoglycin fibers has revealed that they have an extended square lattice with 1494Da connecting rods [6]. The X-ray scattering is dominated by iron atoms at the junctions of these rods. Apart from fibers, hemoglycin also crystallizes into sheets. A flat crystal from an extract of the Sutter's Mill meteorite (sample SM2) has exhibited the typical ladder of diffracted orders associated with this square lattice, extrapolating to a first order spacing of 48-49 Angstroms [5].

In an extended exploration of crystal SM2 the 50micron I24 beam of the Diamond Light Source (UK) was aimed at different positions, revealing an unusual diffraction pattern in the eighth data run. Although there was weak presence at this location of the square lattice diffraction ladder, a much stronger feature was seen for the first time, comprising an exact hexagonally disposed set of short diffraction arcs, spaced on a well-defined ring at 4.085 Angstroms. A frame containing this is shown in Figure S2.2.

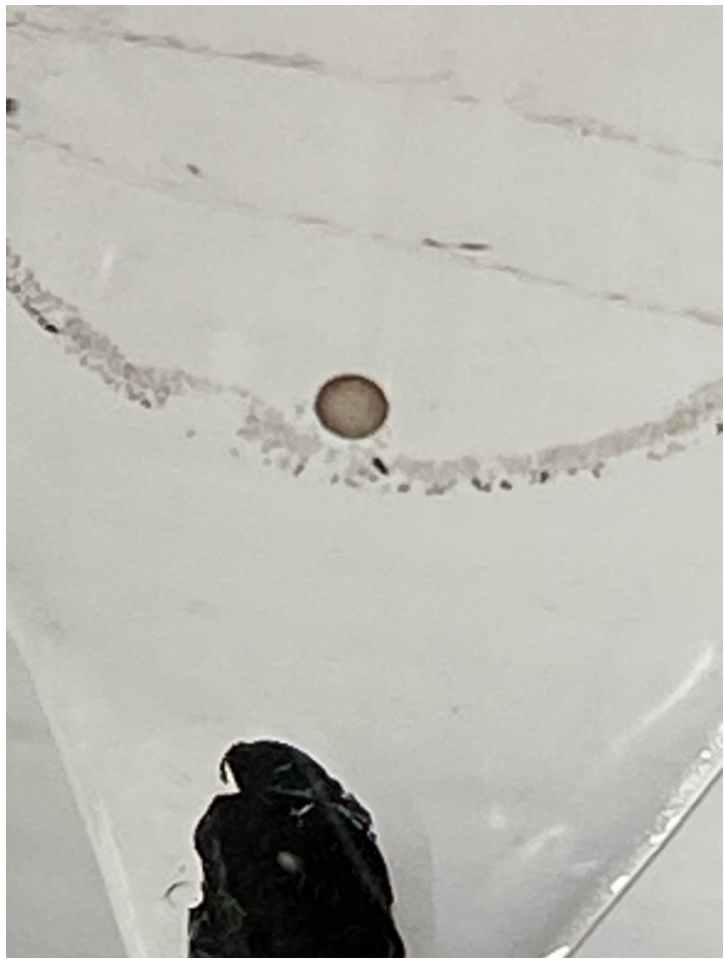

**FIGURE S2.1. Sutter's Mill meteorite particle undergoing Folch extraction in a borosilicate glass V-vial. A spherical dark 533μm diameter vesicle of hemoglycin (image lit from the side) has formed at the interphase between the dense chloroform phase and the polar methanol/water phase. The dark object at the bottom of the vial is the particle of Sutter's Mill meteorite from which the polymer diffuses on solvation. The polymer vesicle resides at the interphase because it is less dense than chloroform but denser than methanol/water due to hydrophobic poly-glycine content.**

This pattern persisted for about 600 images, (i.e. about $60^O$) at each of two $180^O$ opposed locations within a $360^O$ scan. Because it persisted over such a large range of incident angles the object had to have curvature. The full azimuthal distribution of the pattern ruled out the one-dimensional curvature of fiber diffraction, which shows a narrow line of orders [5, 6] leading out from the central beam stop. The possibility of diffraction from a two-dimensional region of curvature was therefore considered. This could not have involved a full sphere in view of the limited persistence across frames of the observed diffraction. However, there could have been a flattened lenticular form if a spherical vesicle had collapsed, and this would show diffraction over a more limited angle range, as illustrated in Figure S2.3.

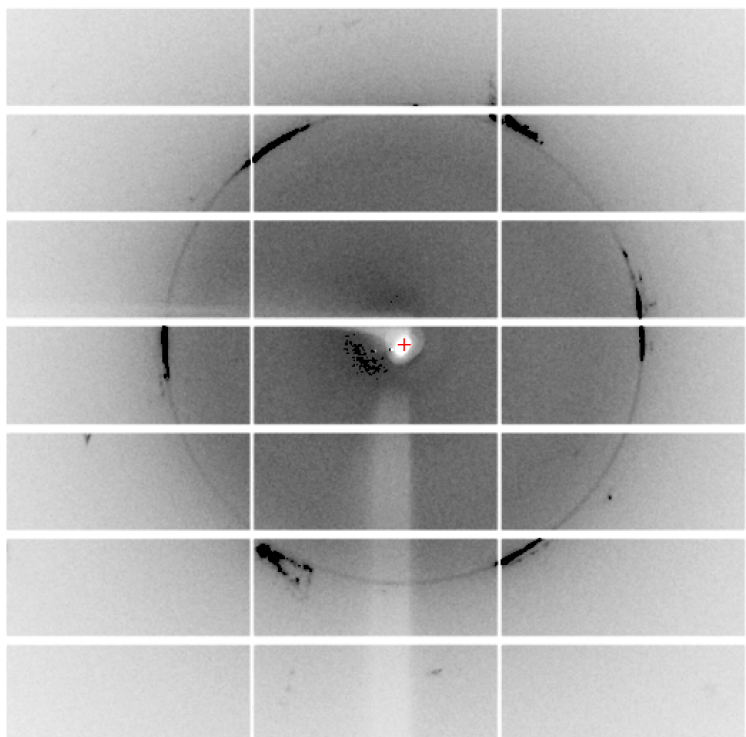

**Figure S2.2 Frame 3100 of Sutter's Mill crystal SM2 at 1.000A, 0.1⁰ oscillation range. The thin circle represents 4.085A spacing.**

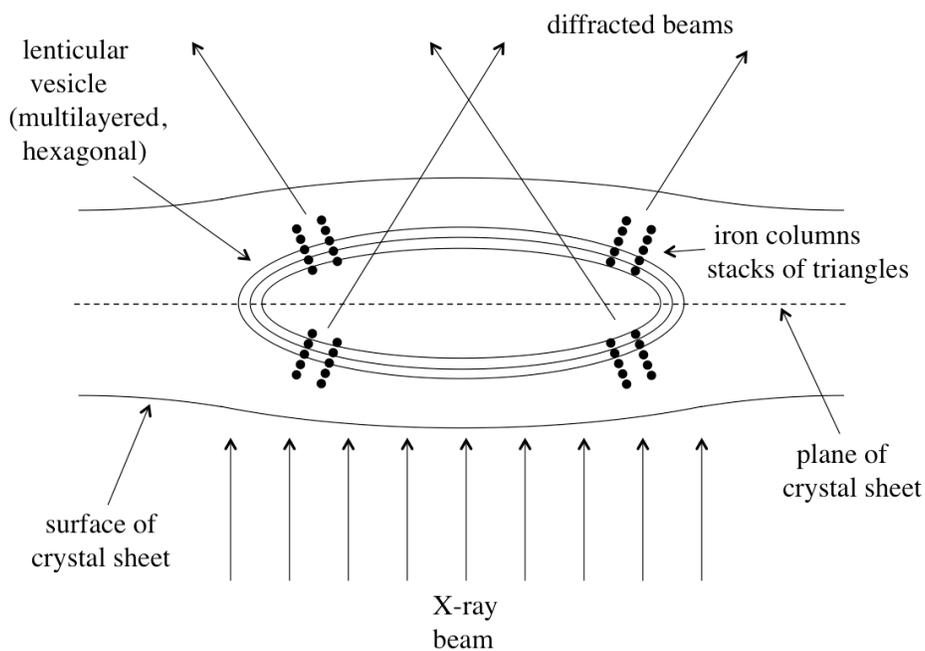

**Figure S2.3. Edge view of a lenticular inclusion in a sheet crystal that could have originated as a spherical vesicle trapped between sheets of the crystal, then dried by evaporation.**

In Figure S2.4 we show the lattice structure that appears to be responsible for the hexagonal pattern of Figure S2.2. This is based upon polymer rods of hemoglycin that are bonded in three-way junctions (also called triskelia) as proposed in [1], able to cover flat or curved areas in combinations of hexagons and pentagons. At each junction there are three iron atoms arranged in a triangle, represented by small dark circles in Figure S2.4 with dotted connections representing either Fe-O-O-Fe or Fe-O-Si-O-Fe. The second of these has been implicated by a mass spectrometry match in the m/z 4641Da entity [1]. However, the first of these, without Si, appears to match the present Sutter's Mill data, being in agreement with a new MMFF simulation that predicts 4.2A separation between Fe atoms in the Fe-O-O-Fe case in good agreement with the observed spacing of 4.085A.

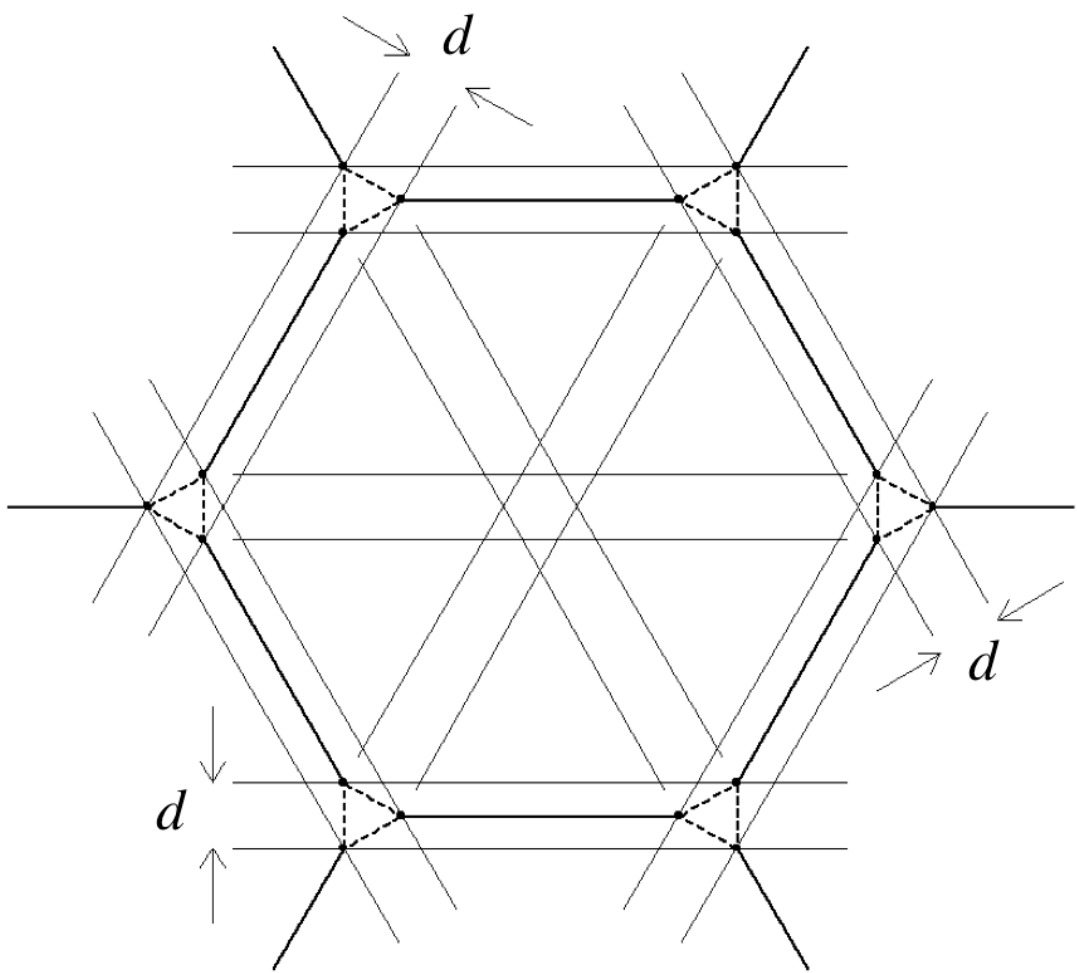

**Figure S2.4. Diffracting planes of separation *d* in a hexagonal lattice with Fe atoms at triangular junctions. The junctions are shown at 2x scale for clarity.**

In Figure S2.4 there are three orientations for scattering, each with the same inter-plane spacing *d*. This structure, when layered over a curved surface will create the opposed pairs of scattering resonances pictured in Figure S2.2, totaling six, as observed. Only a single spacing *d* is possible, not considering the full diagonal of the hexagon at a spacing of about 76 Angstrom that cannot be resolved because it lies above the beam stop limit of about 45A.

In Figure S2.5 we review the hemoglycin 1494Da "core" polymer, here pictured in its triskelion form with an inset showing the triangle of iron atoms at each triskelion junction. The existence of a triskelion had been proposed in [1] to explain the 4641Da entity previously seen via bonding of three of the newly derived 1494Da core units, together with mass spectrometry Na adducts. The version shown does not have Si in the junctions, to match the 4.085A Fe spacing seen here.

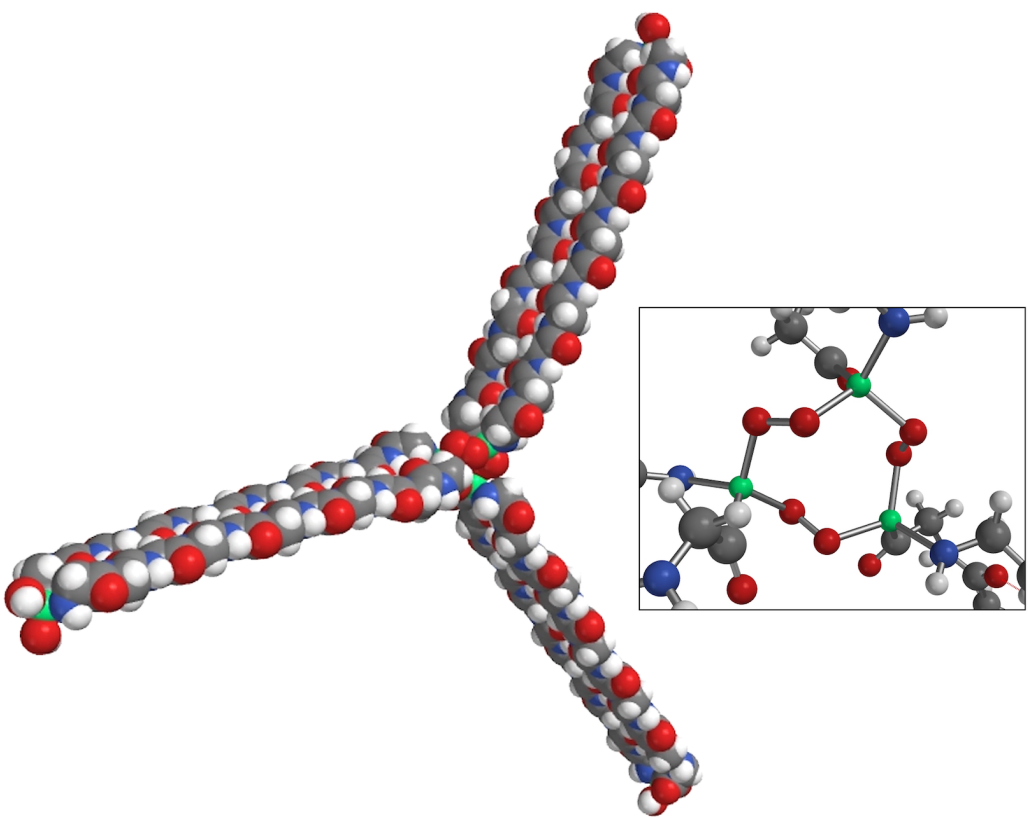

**Figure S2.5. Space-filling representation of the triskelion component of the hexagonal lattice. Fe green, O red, N blue, C black, H white. Insert, the junction region, ball and spoke format.**

We are led to propose that a hemoglycin vesicle of structure different to the bulk sheet crystal has been incorporated and partly flattened, creating a lenticular form as drawn viewed edge-on in Figure S2.3. Although composed of the 1494Da core unit, like the main crystal sheet, this vesicle would have been clad in a hexagonal mesh of triskelia.

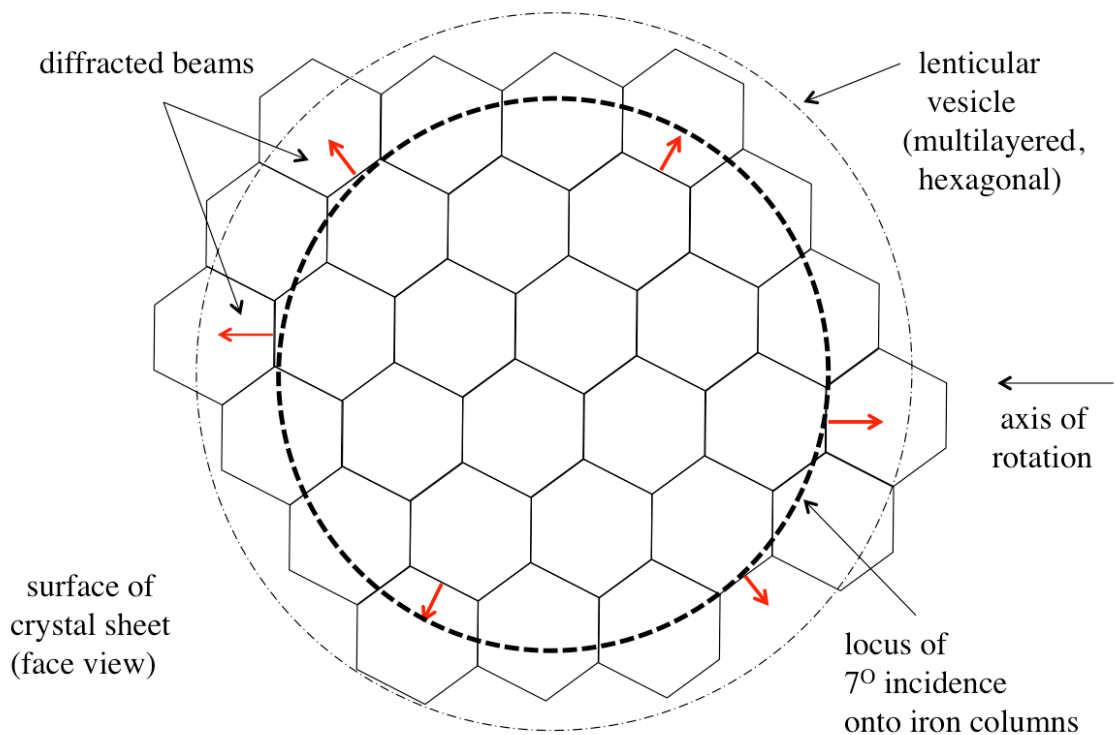

**Figure S2.6. Face view of lenticular hexagonal-surfaced vesicle showing the locus where the 7$^O$ Bragg angle applies, and the six scattering directions. The X-ray beam is perpendicular to the page. Either face of the vesicle can contribute to the scattering.**

**Section S2. Acknowledgements**
The authors thank Diamond Light Source UK for beamtime (proposal MX31420) and Robin Owen, Sofia Jaho and Sam Horrell for hemoglycin crystal X-ray data collection on beamline I24. The Sutter's Mill meteorite sample was kindly provided by Michael E. Zolensky of NASA.